# Controlling thermal emission of phonon by magnetic metasurfaces

X. Zhang[1,2], H. Liu[1*], Z. G. Zhang[1], Q. Wang[1], and S. N. Zhu[1]

[1] *National laboratory of solid state microstructures & school of physics, Collaborative Innovation Center of Advanced Microstructures, Nanjing University, Nanjing 210093, China*

[2] *Shandong Province Key Lab of Laser Polarization and Information, Qufu Normal University, Qufu 273165, China*

Email: liuhui@nju.edu.cn; xzhangqf@mail.qfnu.edu.cn

## Abstract

Our experiment shows that the thermal emission of phonon can be controlled by magnetic resonance (MR) mode in a metasurface (MTS). Through changing the structural parameter of metasurface, the MR wavelength can be tuned to the phonon resonance wavelength. This introduces a strong coupling between phonon and MR, which results in an anticrossing phonon-plasmons mode. In the process, we can manipulate the polarization and angular radiation of thermal emission of phonon. Such metasurface provides a new kind of thermal emission structures for various thermal management applications.

## Introduction

In recent years, the research interest of thermal emission in the infrared wavelength range is growing fast due to its important applications in thermaophotovoltaic (TPV) devices [1], radiative cooling [2,3], incandescent source [4], near-field heat transfer [5], and infrared spectroscopy [6]. Up to now, various structures and systems have been used to control thermal emission, such as gratings [7, 8], nanoantennas [9, 10], photonic crystals [11], surface plasmons [12-14], metamaterials [15-17] and metasurfaces (MTS) [18, 19]. Some properties of thermal emission, such as the emission bandwidth [20], coherent properties [7] and dynamics switching [21, 22] are reported to be controlled in these systems.

In physics, thermal emission is caused by the elementary excitations in materials, such as phonon in polar dielectric [7, 18], exciton in semiconductor [20, 21], and plasmons in metals. Up to now, most of related published papers investigated thermal emission produced by only one elementary excitation. There are few work to explore the simultaneous contribution of different elementary excitations to thermal emission. In the infrared frequencies, phonons, the quantum emitter as the 'fingerprints' of materials characterizing the vibrations of constitution atoms, are the eigen states of materials. And manipulating thermal emission of phonons is important in electronics, optics, bimolecular, and integrated circuits [23-26]. However, the thermal emission of phonon is usually weak and hard to be controlled. On the other hand, plasmons of different metallic nanostructures can be easily controlled through structural designing. In this work, we will show that thermal emission of phonon can be well controlled through phonon-plasmons coupling in magnetic metasurfaces.

Compared with bulk metamaterials, 2-D MTS are more easily fabricated and the optical loss can be suppressed greatly [27-29]. Up to now, MTS has been widely used in anomalous negative refraction [30-32], beam shaping [33-35], surface wave excitation [18, 36, 37], hologram [38, 39], reflection phase [40], mathematical operation [41], polarization controlling [42-47], functional mirrors [48,49], nonlinear devices [50,51], invisible cloaking [52] and subwavelength imaging [53]. Among different metasurface resonance unit designs, metal/insulation/metal (MIM) sandwich resonator has very strong local MR mode and is easily fabricated. MIM resonator has been used in magnetic field enhancement [54], magnetic polariton [55], magnetic plasmon waveguide [56], magnetic nanolaser [57], perfect absorber [58, 59], nonlinear generations [60], negative optical pressure [61] and dipolar response [62].

In this work, the MIM metasurface is used to control the emission of phonons. Using high temperature synthesized amorphous SiO2, we designed and fabricated the Al/SiO2/Al magnetic MTS with Al grating arrayed on the SiO2/Al film (see figure 1 (a)). The MTS grating width is adjusted to produce the strong coupling between MR and phonon inside SiO2. Differ from the precedent studies about the active optical phonons of silica at $\lambda = 10 \mu m$ [63], here, we focus our attention on the relatively inactive optical phonon at $\lambda = 12.5 um$. The absorption and thermal emission spectra were measure by Fourier-transform infrared (FTIR) spectrometer in the wavelength range $11-16 \mu m$. The anticrossing features in the spectra denoted that strong coupling between phonon and magnetic resonance, albeit with big difference in radiation intensity of these two resonant states. Using commercial software FDTD solution and coupled mode theory [64], we analysed the coupling physical mechanism. The theoretical calculations agree with experiments quite well.

**Experiment result**

The Al/SiO2/Al magnetic MTS were fabricated on Si wafer. In fabrication, Al film of thickness 150 $nm$ was deposited on pre-cleaned Si wafer by electron beam evaporation. Amorphous silica films (500 $nm$) were synthesized by plasma enhanced chemical evaporation (PECVD) at temperature 300℃ on the Al film, and subsequent in situ annealing was performed for 10 hours. An Al grating is fabricated on the amorphous SiO2 films by ultraviolent photolithography and subsequent lift-off. The schematic view in Fig.1 (a) exhibits both structure and material information. In experiment, the period and height of grating was fixed as $\Lambda = 6.5 \mu m$ and $h = 0.05 \mu m$, the width d can be changed through tuning the exposure time in the course of photolithography. Fig.1 (b) show the morphology characterized by a commercial scanning electron microscopy (SEM) of sample with width $d = 2.8 \mu m$. The total thickness of MTS is 0.7 micron which is much smaller than the working wavelength range $11-16 \mu m$.

We measured thermal emission and absorption of sample with FTIR spectrometer. The scanning wavelength range is $11-16 \mu m$. At the same time, we also used FDTD method to calculate the absorption of structures and compare the results with experiment. Firstly, we investigated the MR and phonon separately without coupling effect between them. Figure 2 (a) gives the thermal emission and absorption of SiO$_2$/Al. Without grating, there is no MR and only phonon is found. Both the measured absorption (blue triangle symbolic line) and thermal emission (red brackets symbolic line) spectra show one faint and broad peak at $12.5 um$, along with the measured ones was the simulated absorption spectrum by FDTD method as denoted by the black triangle symbolic line in figure 2 (a). The data of SiO$_2$ were referred to ref [65]. The agreement between measured and calculated absorption spectra consolidated our material research base and method reliability. Figure 2 (b) gives the calculated absorption spectrum of MTS, where the refraction index of SiO$_2$ n=1.47 was used in simulation. In the simulation, the MTS grating periodic is $\Lambda = 6.5 \mu m$ and width $d = 3 \mu m$ under TM polarization (electric field perpendicular to the grating), as we can see, only MR resonance peak is found at $12.5 um$. Compared with phonon, the absorption of MR is much stronger which is attributed to the strong magnetic resonance mode [54-61]. Here, the MR mode can be regarded as a kind of plasmonic cavity mode.

For the MTS given in figure 1, its thermal emission shows very strong polarization dependence. Figure 2 (c) gives TE polarized emission (E field parallel to grating strip). Both the thermal emission and absorption spectra of the MTS in Fig.2 (c) show the similar peculiarity as phonon at $12.5 um$ given in figure 2 (a). Here, MR cannot be excited, and only phonon contribute to the emission and absorption. Therefore, these curves are very like those in figure 2 (a). The calculation agree with experiment quite well.

On the other hand, for TM polarized emission of MTS (E field perpendicular to grating strip) in figure 2 (d), it is quite different from TE emission in figure 2 (a). Here, phonon resonance can be still found at the wavelength $12.5 \mu m$. At the same time, the MR can be excited at the same wavelength. This makes MR and phonon overlapped with each other. As a result, the strong coupling occurs between MR and phonon, which will produce two resonance peaks in the curve. In figure 2 (d), two prominent peaks are obtained at wavelength $11.8 \mu m$ and $13.29 \mu m$, which lie on the two sides of $12.5 \mu m$, and are different from peaks of the phonon (figure 2 (a)) and the MR (figure 2 (b)). In simulations, with including phonon of $SiO_2$ in the program, the simulated absorption spectra of sample with periodic $\Lambda = 6.5 \mu m$ and width $d = 3 \mu m$ also features two pronounced peaks as denoted by the black triangle symbolic line given in figure 2 (d). The simulated and measured absorption spectra agree well, which are both in accordance with the thermal emission spectrum. In the following, we will provide a theory model to explain the coupling mechanism.

The absorption and emission polarization dependence of sample can be described by the ratio of TM over TE: $P=I_{TM}/I_{TE}$. For $SiO_2$/Al film, there is no polarization dependence for phonon mode and P=1. While, for MTS, strong anisotropic property is demonstrated for both theoretical and experimental results as shown in figure 3 (experimental result: red line, and theoretical result: blue line). We can see P is very large for the wavelength around the two resonance peaks, while it will approximate unit when the wavelength is far away from the peaks. It means that the polarization dependence is caused by the strong resonance mode of MTS. In the figure, P amounted to forty-fold in calculations, and only about five-fold in experiment. This large discrepancy between theory and experiment is caused by the defects of samples and the aberrations in adjusting the polarizations. Despite all this, the modified anisotropic thermal emission of phonon was still clearly discerned. Beside polarization dependence, the angular dependence of emission and absorption is also changed by MTS. The figure 4 (a) and (c) give the results of $SiO_2$/Al, in which only phonon exists. As given in ref [63], phonon near $12.5 \mu m$ includes both LO and TO pair mode in the range 1160-1200 cm$^{-1}$. Due to Berreman effect of LO phonon in thin film [66], the interaction between phonon and light will be enhanced by increasing the angle. Then, as shown in figure 4 (a) and (c), the absorption and thermal emission of SiO$_2$/Al film became stronger at larger emission or incident angle. While for MTS in figure 4 (b) and (d), both the two resonance modes obtain their maximum absorption and emission efficiency at normal angle. For the mode at wavelength $11.8 \mu m$, the absorption and emission do not change much with angle. While, for the mode at wavelength $13.29 \mu m$, the emission and absorption is reduced greatly around 50 degree. This is caused by strong Bragg scattering at this angle.

In above discussion, we know that the coupling between phonon and MR will produce two resonance peaks. In physics, it is like the normal mode splitting of atoms in cavity [67-69]. By putting atom into high-finesse optical cavity, and tuning the parameters of the cavity, the spontaneous emission spectra of the atom featured anti-crossing line-shape, which was attributed to the strong coupling occurred between the atom and the cavity. Actually, such mode splitting was also possibly for magnetic polariton structures [55]. In our system, if the MR can be regarded as a cavity and phonon as an atom, the two peaks can be seen as the result of mode splitting. Then if we tune the MR by changing structure parameter, it is possible to produce anti-crossing line-shape near the phonon resonance. For the MTS in figure 1 (a), its MR resonance wavelength is dependent on the grating width d. Then we can tune MR through changing d. FDTD simulations were performed with varied grating width d of MTS. Fig.5 (a) depicted the simulated absorption spectra versus the photon energy and the grating width d, which was changed from 2.6 to 3.6 $\mu m$. Pronounced anticrossing behaviour was demonstrated in the absorption spectra, indicating the presence of the strong coupling between MR and the phonon. In experiment, a series of samples with varied width d and fixed period $\Lambda = 6.5 \mu m$ were synthesized through changing the exposure time in the process of photolithography. The measured thermal emission spectra of the varied width d grating were shown in figure 5 (b), where the measurements were performed in normal angle and TM polarization. The pronounced anticrossing behaviour near phonon energy $\omega = 0.1ev$ ( $\lambda = 12.5\mu m$ ) was also demonstrated in both figure 5 (a) and (b). The excellent agreement between simulation and experiment render the strong coupling convincible in this system.

**Theory**

In this part, a coupled mode theory [64] is established to describe the physical mechanism of the strong coupling between the MR and the phonon. Firstly, let's recur to the common equations that describe two coupled oscillators as:

$$\begin{aligned}\ddot{\vec{\mu}}_1 + \gamma_1 \dot{\vec{\mu}}_1 + (\omega_0 + \delta)^2 \vec{\mu}_1 + \kappa \vec{\mu}_2 &= g_1 A e^{i\omega t} \\ \ddot{\vec{\mu}}_2 + \gamma_2 \dot{\vec{\mu}}_2 + \omega_0^2 \vec{\mu}_2 - \kappa \vec{\mu}_1 &= g_2 A e^{i\omega t}\end{aligned} \quad (1)$$

where $\mu_1, \mu_2$ are the displacements of oscillators under excitation, respectively, and $g_1, g_2$ are coupling factor between oscillators and incident field, $\kappa$ is the coupling factor between the two oscillators [64]. $\delta$ is the resonance frequency difference between $\mu_1$ and $\mu_2$. $\gamma_1$ and $\gamma_2$ are dissipation losses of oscillators $\mu_1, \mu_2$, respectively. Here, we choose oscillator $\mu_2$ as the phonon, with inherent resonant frequency $\omega_0$. And the MR oscillator $\mu_1$ can be manipulated through changing the parameters of the MTS. The Hamiltonian of this two oscillators system can be presented as:

$$H = \begin{pmatrix} \delta + \omega_0 - i\gamma_1 & \kappa \\ \kappa & \omega_0 - i\gamma_2 \end{pmatrix} \quad (2)$$

Due to the coupling term in Hamiltonian, the mode splitting happens and the two eigen frequencies can be obtained as:

$$\omega_\pm = \omega_0 + \left(\delta - i(\gamma_1 - \gamma_2) \pm \sqrt{(\delta - i(\gamma_1 - \gamma_2))^2 + 4\kappa^2}\right)/2 \quad (3)$$

$$\kappa = \pm \frac{i\delta - (\gamma_1 - \gamma_2)}{2} \quad (4)$$

When the MR resonant frequency is near the phonon frequency $\omega = 0.1ev$, namely, $\delta \approx 0$, the strong coupling may occur if $\kappa \gg (\gamma_1 - \gamma_2)^2$ [67-69], and the coupled system has two eigen energies, the normal mode splits into two modes, and thus two peaks will be found in absorption spectra.

Here, the mode splitting is like the Rabi splitting of atom in cavity and the Rabi frequency is [67-69]:

$$\hbar\Omega_{rabi} = 2\sqrt{(\hbar\kappa)^2 - \frac{1}{4}(\hbar\gamma_1 - \hbar\gamma_2)^2} \tag{5}$$

Based on the equation (3), the eigen frequencies of the MTS with varied grating width were numerically calculated, where the parameters of SiO$_2$ phonon $\gamma_2 = 0.65254(ev)$ were fitted from the Lorentz formulation. Figure 6 displays the eigen-wavelength versus the grating width d from coupled mode theory (the blanket triangles) and FDTD simulation (the solid red dots). The green dotted line in figure 6 indicates the phonon resonance wavelength, and the purple dotted line depicts the MR wavelength varied with the grating width, the error bar gave out the difference value between them and the experiment results. It is evident that the coupled mode theory calculation, the FDTD simulation and the experiment agrees quite well. All the three spectra feature anticrossing characterization. The narrowest frequency is located rightly at the phonon resonant frequency, $\omega = 0.1ev$, with $\gamma_1 = 1.06896(ev)$, and thus the detuning $\delta = 0$. Substitute these parameters into formula (4), the coupling coefficient can be obtained as $\kappa = 0.65265(ev)$. Clearly, $\kappa^2 \gg (\gamma_1 - \gamma_2)^2$, satisfies the strong coupling condition [67-69]. Substituting these parameters into Rabi splitting energy band formula (5) $\hbar\Omega_{rabi} = 2\sqrt{(\hbar\kappa)^2 - \frac{1}{4}(\hbar\gamma_1 - \hbar\gamma_2)^2}$ [67, 70, 71], and we get the Rabi splitting band gap in the anisotropic magnetic MTS as $\hbar\Omega_{rabi} = 1.237(ev)$. Hitherto, from the excellent agreement between simulation and experiment results, we conclude that the thermal emission of SiO$_2$ phonon at $\omega = 0.1ev$ can be controlled by MTS.

## Summary

In this work, we use MTS to control the thermal emission of phonon. The emission peak, polarization and radiation angle can be well manipulated in the process. A coupled mode theory is established to calculate the mode splitting and anti-crossing effect, which agree with experiment well. In this work, we only consider the phonon inside $SiO_2$. Actually, this method can be used to any other materials with different phonon wavelength as the MR can be flexibly tuned to any wavelength. If the fabrication is improved and the MTS have larger resonance Q factor, the plasmon-phonon coupling can be further enhanced. In the future, it can be anticipated that MTS will have many other interesting applications in thermal emission devices.

## Acknowledgments

This work was supported by the National Natural Science Foundation of China (No's 11321063, 61425018 and 11374151), the National Key Projects for Basic Researches of China (No. 2012CB933501 and 2012CB921500).

## Author contributions statement

X. Zhang and Z. G. Zhang performed the sample fabrication and measurement, X. Zhang and Q. Wang performed numerical simulations. H. Liu and Q. Wang provided helpful discussions. X. Zhang and H. Liu wrote the manuscript. H. Liu and S.N. Zhu initiated the program and directed the research.

## Additional information

**Competing financial interests**: The authors declare no competing financial interests.

# Figure captions

Figure 1 (color online) the sketch plot (a) and SEM top view (b) of sandwiched structure. In experiments, the periodic and height are fixed with $\Lambda = 6.5 \mu m$ and $h = 0.05 \mu m$, d is changed from 2.6 to 3.6 $\mu m$.

Figure 2 emission (red bracket) spectra and absorption spectra of experiment (blue triangle) and of calculation (black triangle) for $SiO_2/Al$ film (a), for $Al/SiO_2/Al$ MTS under TE polarization (c) and for $Al/SiO_2/Al$ MTS under TM (d). (b) Simulated absorption of $Al/SiO_2/Al$ MTS without $SiO_2$ phonons under TM.

Figure 3. Anisotropic ability P versus wavelength for sandwiched MTS with experiment (the red line) and calculation (the blue line).

Figure 4 simulated absorption (a) (b) and measured thermal emission (c) (d) spectra of $SiO_2$ film (a) (c) and $Al/SiO_2/Al$ sandwiched MTS (b) (d), respectively.

Figure 5. Simulated absorption (a) and measured emission (b) spectra versus photon energy and grating width d.

Figure 6. Absorption peaks extracted From the FDTD simulation (red dotted line) and the numerical (blue blanket triangle) results. The error bar give out the deviative of them from experiment.

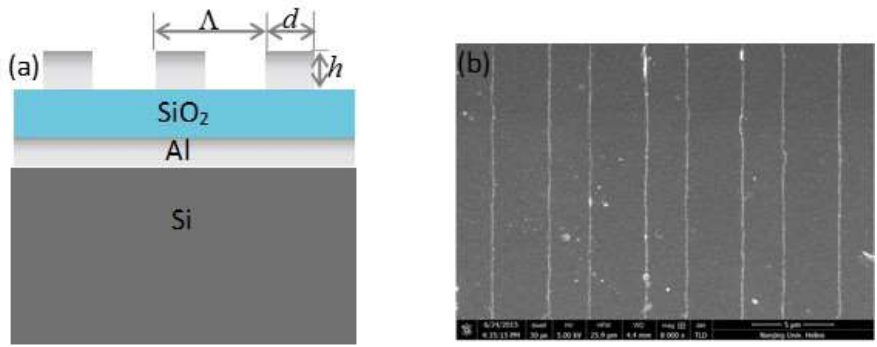

Figure 1

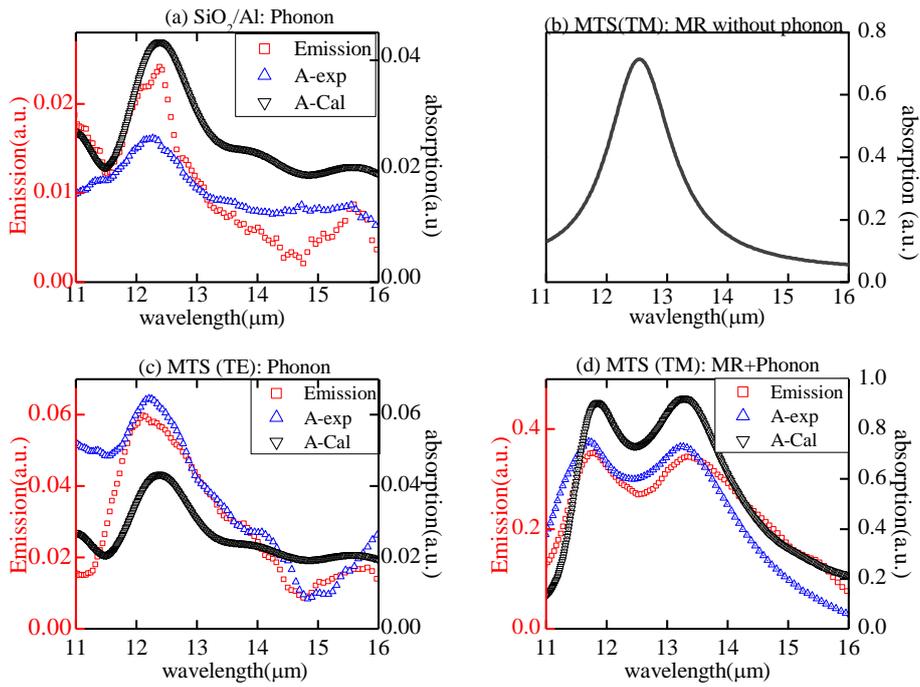

Figure 2

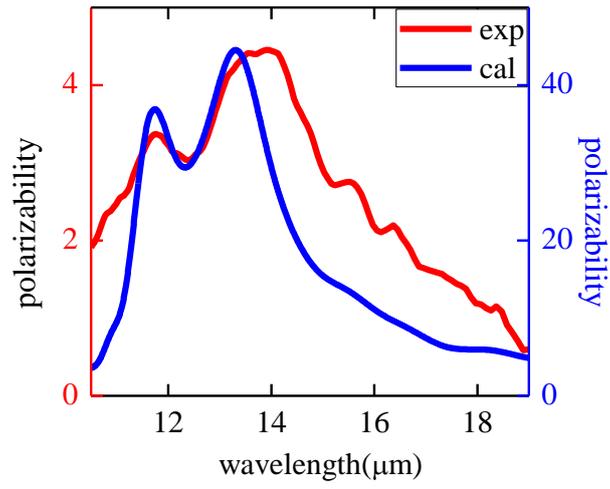

Figure 3

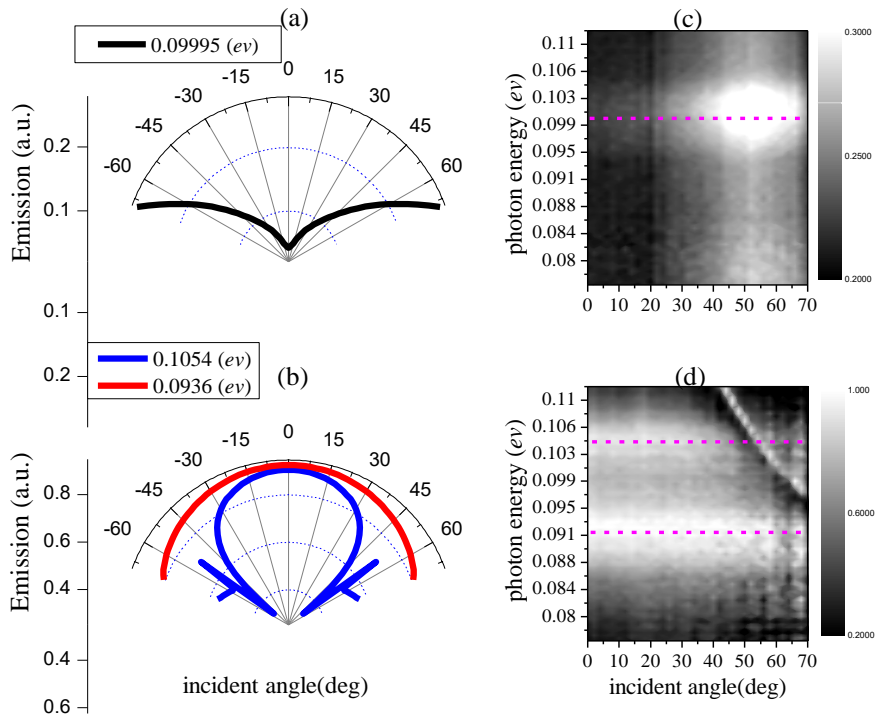

Figure 4

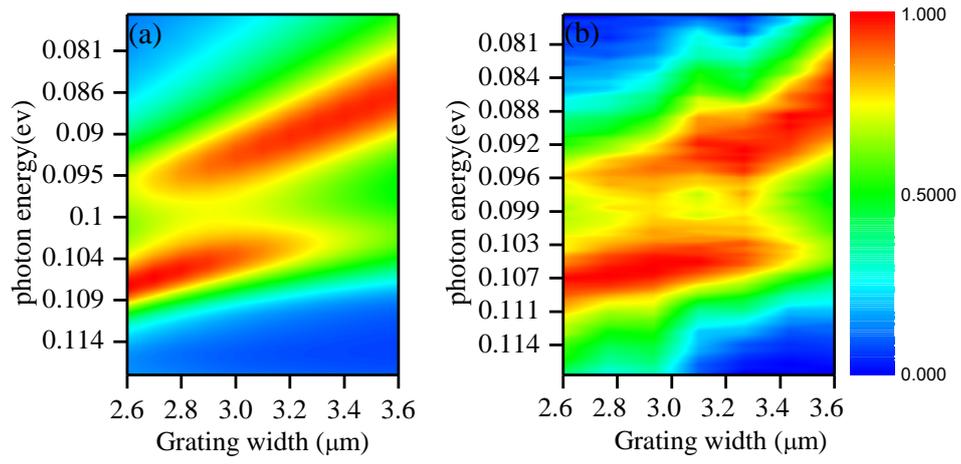

Figure 5

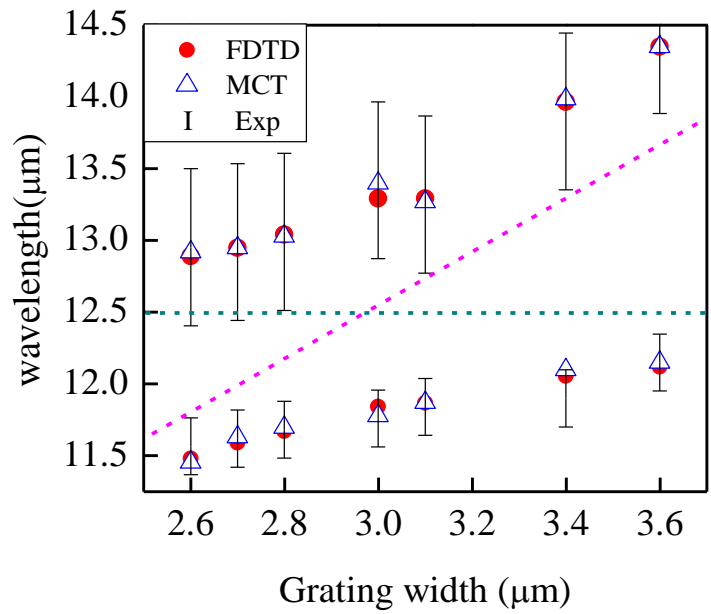

Figure 6